\documentclass{article}
\IfFileExists{icml2026.sty}{%
  \usepackage{microtype}
  \usepackage{graphicx}
  \usepackage{booktabs}
  \usepackage{hyperref}
  
  \usepackage[preprint]{icml2026}
  \newcommand{\ICMLKIT}{1}
}{%
  \usepackage[letterpaper,margin=0.75in,columnsep=0.25in]{geometry}
  \usepackage{times}
  \usepackage{microtype}
  \usepackage{graphicx}
  \usepackage{booktabs}
  \usepackage[numbers,sort&compress]{natbib}
  \usepackage{hyperref}
  \twocolumn
}
\usepackage{amsmath}
\usepackage{xurl}   

\usepackage{xcolor}

\newcommand{\gain}{\Delta_{\mathrm{CK}}}

\begin{document}

\ifdefined\ICMLKIT
\icmltitlerunning{Does Agreement with Clonal Fate Measure Inferred Cell Dynamics?}
\twocolumn[
\icmltitle{Does Agreement with Clonal Fate Measure Inferred Cell Dynamics?}
\begin{icmlauthorlist}
\icmlauthor{Xuanrui Huang}{nus}
\end{icmlauthorlist}
\icmlaffiliation{nus}{National University of Singapore, Singapore}
\icmlcorrespondingauthor{Xuanrui Huang}{huangxuanrui@u.nus.edu}
\vskip 0.2in
{\begin{center}
{\includegraphics[width=\textwidth]{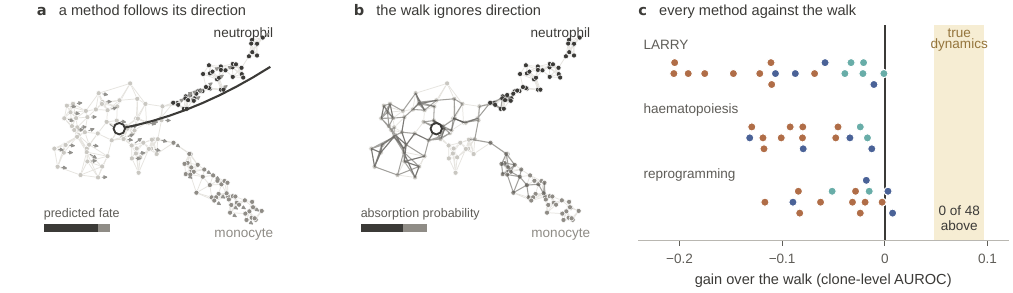}}
\captionof{figure}{What is compared, and what comes out. (a) A dynamical method predicts the fate
of an early cell (open circle) by following its inferred direction. (b) The opponent: on the same
neighbour graph and terminal states, a random walk without direction, whose absorption probabilities
are its fate prediction. (a) and (b) are schematic. (c) Gain of every method over the walk on three
lineage-traced data sets (clone-level AUROC), coloured by method family (couplings, learned
generative models, CellRank kernels; Table~\ref{tab:setup}); 36 are significantly below the walk.
Gold: the gain that true dynamics add in simulation at these data sets' level. None of
the 48 method--data set pairs is above the walk.}
\label{fig:concept}
\end{center}}
\vskip 0.15in
]
\printAffiliationsAndNotice{}
\else
\title{Does Agreement with Clonal Fate Measure Inferred Cell Dynamics?}
\author{Xuanrui Huang}
\date{}
\maketitle
\fi

\begin{abstract}
Methods that infer cell dynamics from single-cell snapshots, from RNA velocity to optimal transport,
flow matching and neural differential equations, often support their dynamics with one test: predict
the fate of early cells and compare it with the fate their sister cells reach, as recorded by heritable
barcodes. We ask whether this test measures dynamics. As an opponent we use a random walk that ignores
direction, on the same neighbour graph and towards the same terminal states. Of 48 method--data set
pairs (18 methods, three lineage-traced data sets), none beats it: 36 score below it and 12 cannot
be separated from it. On the 335 test cells of the published LARRY benchmark, the walk reaches
Pearson $r = 0.33$--$0.42$ across graph settings, the range of the best published dynamical model
($0.40$--$0.41$) and above the three methods the benchmark was introduced with ($0.15$--$0.23$). The score
comes from how close an early cell already is to each mature population, in its own expression and
through later cells on the shared graph; the walk reads both. We therefore score each method by its
gain over the walk, with a paired interval over clones and a minimum detectable difference. A
simulation with known dynamics calibrates this score: at the level of the real data sets, dynamics
that carry fate would add $0.05$--$0.10$ AUROC, large enough for the test to detect. Scored this
way, one method rises above the walk, by a small margin and mostly through one rare fate
(StationaryOT over six fates on LARRY, $+0.012$). What the learned models do
learn shows at the level of populations: on a held-out time point, four of them beat a control
without dynamics. Code: \url{https://github.com/xuanruihuang/fatescore}.
\end{abstract}

\section{Introduction}

A growing family of methods infers how cells change over time from sequencing snapshots. Machine
learning contributes potential-based generative models such as PRESCIENT~\citep{yeo2021}, flow
matching and Schr\"odinger bridges~\citep{tong2024otcfm,tong2024sf2m}, and neural ordinary or
stochastic differential equations fitted across time points~\citep{huguet2022mioflow,zhang2024scnode};
biology contributes RNA velocity~\citep{lamanno2018,bergen2020} and the Markov chains of
CellRank~\citep{lange2022,weiler2024}.

A dynamical model is a rule of change: from the state of a cell now, where it moves next, how fast,
and driven by what. What it offers beyond a map of cell states is what such a map cannot give: the
direction and timing of change, states at times that were not measured, and the response to an
intervention. A snapshot holds no trajectories, so a method has to show that what it learned is
dynamics. The evidence cited most often is clonal fate. Lineage tracing such as LARRY~\citep{weinreb2020} marks
cells with heritable barcodes, so an early cell can be matched with the cell types its sisters
become. A method predicts the fate of early cells from its inferred dynamics, the prediction is
compared with the sisters' fate, and a higher agreement is read as more accurate dynamics. The test
set Weinreb et al.\ built on LARRY was adopted by PRESCIENT, and the LARRY fate task has since
supported TIGON~\citep{sha2024tigon}, MultistageOT~\citep{tronstad2025}, Chreode~\citep{qiu2026chreode},
LIFT~\citep{yang2026lift} and LineageFlow~\citep{liu2026lineageflow}, each under its own protocol.
MultistageOT and LineageFlow report baselines without dynamics (distance to the mature cell types;
the empirical class prior), and LineageFlow reports intervals over clones; none compares with a
predictor without direction on the method's own graph and terminal states.

The reading rests on an unchecked assumption: that a high score comes from the
dynamics. If fate is already largely written in the present state of a cell, anything that connects
early cells to mature cells on a graph can score high, with or without a direction: the benchmark
would reward a shortcut~\citep{geirhos2020shortcut}. We therefore
give every score an opponent: CellRank's connectivity kernel, a random walk on the same neighbour
graph towards the same terminal states that uses only symmetric similarities (Fig.~\ref{fig:concept}). We ask four questions.
Do the methods exceed it? If the score does not come from the dynamics, where does it come from?
Would the test see true dynamics if they were there? And which tests do see them?

\paragraph{Contributions.}
\begin{itemize}
\item \emph{A failure mode, at scale.} Across three lineage-traced data sets and 18 methods (optimal
transport, flow matching, a potential-based model, six neural dynamical models, five CellRank
kernels), none of 48 pairs exceeds the walk without direction. On the published benchmark's own
test cells the walk is level with the best published result (Fig.~\ref{fig:published}). Distribution fit and fate
accuracy rank the neural models differently.
\item \emph{Where the score comes from.} Fate is largely readable from how close an early cell is to
each mature population: from its own expression (a classifier on day-2 expression and the sorted
population reaches AUROC $0.87$ on LARRY) and through later cells on the graph. A simulation separates how much fate
the present state carries from how much the dynamics carry; the walk reads only the first. Random error in the dynamics only erodes their
gain towards zero; falling below the walk requires directions that favour the wrong fate.
\item \emph{What the raw score misses.} Reported raw, agreement with clonal fate credits a method
with what the walk already has. Scored against the walk, one method rises above it, slightly
(StationaryOT over six fates, $+0.012$);
what the learned models do learn shows on a held-out time point, where four of them beat a control
without dynamics on LARRY, level with an optimal-transport interpolation.
\item \emph{A calibrated score and a tool.} We report each method as a gain over the walk with a
paired interval over clones and a minimum detectable difference; positive controls show that the
gain of true dynamics ($0.05$--$0.10$) is within reach, and that a learned model run through the
same pipeline rises above the walk where the graph misleads. Code and one command per figure are
released.
\end{itemize}

\paragraph{Related evaluations.} Existing evaluations rank methods against each other; this one
asks whether the ranking test measures dynamics at all. Trajectory-inference benchmarks score
ordering and topology~\citep{saelens2019}, scTimeBench scores distribution fit and cell-type
paths~\citep{sctimebench}, and a recent RNA-velocity benchmark scores against labelling, the cell
cycle and lineage time labels~\citep{wu2026}. Each ranks methods; none tests whether its score
requires dynamics. A control without dynamics on the same graph answers that question, and for
clonal fate it wins. CoSpar observed that smoothing over the cell-similarity graph improves fate
prediction from state alone and built a method on it~\citep{wang2022cospar}; we turn that graph into
the opponent every method must beat. LineageOT normalises its lineage ground truth by a global
independent coupling~\citep{forrow2021}, which leaves the graph's information on the method's side.
Methods that use the clones in training, such as Super-OT~\citep{prasad2020superot}, fall outside
this test. Newer models with growth report distribution fit alone, even on lineage-traced
data~\citep{wang2025vgfm}; fit and fate rank models differently (Section~\ref{sec:eighteen}).
Critiques of RNA velocity question the velocity estimates~\citep{bergen2021,gorin2022,zheng2023};
we question the test, and the finding extends to optimal transport, flow matching and neural
differential equations. CellRank introduced both kernels~\citep{lange2022,weiler2024}; we provide the
comparison between them against a lineage ground truth. Simple baselines that match deep models are
known in perturbation prediction~\citep{ahlmanneltze2025,wenteler2025}; as in~\citet{needles2026},
we go on to the cause and to a calibrated score.

\section{The test and its opponent}

\paragraph{Tasks.} In each data set we take the early cells whose clone can be traced to later
time points (the task cells). The target is the share of one fate among the clone's mature sisters: neutrophil versus
monocyte in LARRY (567 clones); monocyte-dominated clones versus others in CellTag-multi
haematopoiesis~\citep{jindal2024} (667 clones); reprogramming-dominated clones versus others in
CellTag-multi reprogramming (304 clones). The main score is the AUROC over clones of the method's
predicted fate probability; on the published LARRY test set we also use the Pearson $r$ it was
introduced with.

\paragraph{Opponent.} CellRank's connectivity kernel (CK; below, \emph{the walk}) uses the same
neighbour graph and terminal states
as the method and turns symmetric similarities into a random walk; absorption probabilities into the
terminal states are its fate prediction. It contains no direction. We build it at 18 graph settings
(neighbours 15/30/60, principal components 15/30/50, with and without batch correction); 30
neighbours and 30 components with correction is the reference, and the others give a band. Terminal states are the published
annotations of mature cells, not states found by an algorithm. Learned models use no graph; their
simulated endpoints are classified against the same annotated cells of the last day, so every
method and the walk are scored against the same fates and see the same later cells.

\paragraph{Gain and interval.} For a method $m$, $\gain(m) = \mathrm{AUROC}(m) -
\mathrm{AUROC}(\mathrm{CK})$ on the same clones. A paired bootstrap over clones (2,000 resamples)
gives a 95\% interval; the minimum detectable difference is $0.714$ times the interval width, the
difference a two-sided 5\% test detects with 80\% power (Appendix~\ref{app:tool}). A method is \emph{above} if the interval
lies above zero, \emph{below} if it lies below, and \emph{not separated} otherwise.

\begin{table}[t]
\centering
\small
\setlength{\tabcolsep}{3pt}
\caption{Data sets and methods.}
\label{tab:setup}
\begin{tabular}{@{}p{0.30\columnwidth}p{0.66\columnwidth}@{}}
\toprule
\multicolumn{2}{@{}l}{\emph{Lineage-traced data sets (clonal fate)}} \\
LARRY & days 2, 4, 6; 567 clones, 293 neutrophil-dominated~\citep{weinreb2020} \\
CellTag-multi haem. & days 2.5, 5; 667 clones, 533 monocyte-dominated~\citep{jindal2024} \\
CellTag-multi reprog. & days 3, 12, 21; 304 clones, 112 reprogramming~\citep{jindal2024} \\
\midrule
\multicolumn{2}{@{}l}{\emph{Metabolic labelling (Appendix~\ref{app:labelling})}} \\
scNT-seq & haematopoiesis~\citep{qiu2022dynamo} \\
scEU-seq & RPE1 with FUCCI; intestinal organoid~\citep{battich2020} \\
\midrule
\multicolumn{2}{@{}l}{\emph{Methods, in three families (the colours of every figure)}} \\
Couplings & Waddington-OT~\citep{schiebinger2019}, moscot~\citep{klein2025moscot},
StationaryOT~\citep{zhang2021statot}, CoSpar (state only)~\citep{wang2022cospar} \\
Learned generative models & PRESCIENT~\citep{yeo2021}; flow matching: OT-CFM~\citep{tong2024otcfm},
SB-CFM~\citep{tong2024sf2m}; neural ODE/SDE: scNODE~\citep{zhang2024scnode}, scIMF~\citep{scimf},
PI-SDE~\citep{pisde}, MIOFlow~\citep{huguet2022mioflow} (run by scTimeBench~\citep{sctimebench}),
USB~\citep{usb2026}, PACE~\citep{pace2026} \\
CellRank kernels & velocity (dynamical, steady state)~\citep{bergen2020}, velocity +
connectivity, pseudotime, real-time~\citep{weiler2024} \\
\midrule
Opponent & the walk: CellRank's connectivity kernel, 18 graph settings \\
\bottomrule
\end{tabular}
\end{table}

\paragraph{Running the methods.} Each method uses the settings of its paper or tutorial. Training is
judged only by distribution fit: the Wasserstein distance between simulated and observed cells,
next to the distance between two batches of the same day (floor) and to leaving the cells in place.
Settings were chosen by fit alone, never by clonal scores. The four methods wrapped by
scTimeBench (scNODE, scIMF, PI-SDE, MIOFlow) are trained by its runner. Our adapter then carries
each early cell forward with the trained model and classifies the endpoint by its 20 nearest
observed cells of the last day, as PRESCIENT does. Versions and deviations are listed in
Appendix~\ref{app:deviations}.

\section{Do the methods beat the walk?}

Not on the task as published: neither on the benchmark's own test cells nor on three data sets with
18 methods. Scored over six fates, one method edges above the walk.

\subsection{On the published benchmark}

\begin{figure}[t]
\centering
\includegraphics[width=\columnwidth]{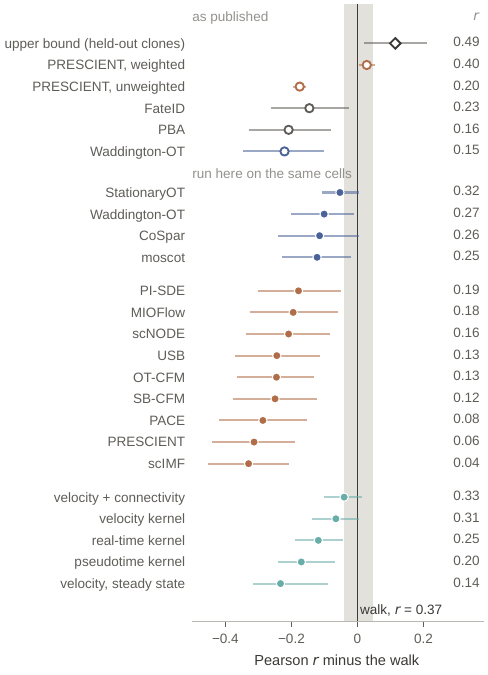}
\caption{The LARRY fate benchmark on its own test cells. Pearson $r$ between predicted and observed
neutrophil share on the 335 test cells, as a difference from the walk (dots, 95\%
paired bootstrap over 298 clones). Grey band: the walk at 18 graph settings. Top block, open
markers: published values (diamond: the held-out clonal estimate, an upper bound; PRESCIENT with
and without proliferation weights: mean $\pm$ SD over five seeds, not paired). Waddington-OT appears
twice: Weinreb et al.'s predictions and our run. Bottom block, filled: our runs, coloured by method family as in
Fig.~\ref{fig:concept}; PRESCIENT here is the run under the protocol shared by all methods, its most
faithful reproduction ($0.16$) is in Appendix~\ref{app:deviations}. Numbers on the right: absolute $r$.}
\label{fig:published}
\end{figure}

We rebuilt the 335 day-2 test cells of Weinreb et al.\ from the files they shared and reproduced the
scores PRESCIENT reports for FateID ($r = 0.225$), WOT ($0.150$) and the held-out clonal upper
bound ($0.485$ versus $0.487$). All 335 cells (298 clones) are in our processed object, matched by
barcode, and their clonal shares equal the published targets on every cell.

On these cells the walk reaches $r = 0.371$ at the reference setting and
$0.33$--$0.42$ over its settings (Fig.~\ref{fig:published}). The best published dynamical model, PRESCIENT, lies inside this band:
$0.399 \pm 0.025$ with proliferation weights estimated from cell-cycle and apoptosis genes, and
$0.407 \pm 0.019$ at its best setting. Without the weights the same paper reports $0.196 \pm 0.020$,
below the walk. The step from $0.20$ to $0.40$ comes from weights computed from each cell's present
expression, not from the inferred motion. The
three methods the benchmark was introduced with, FateID, PBA and WOT, reach $0.23$, $0.16$ and
$0.15$, all significantly below the walk; the upper bound is $0.49$. All 18 methods we ran under
the same protocol are below the walk or not separated from it (StationaryOT, CoSpar and the
velocity kernels, $0.26$--$0.33$). Our first PRESCIENT run (Fig.~\ref{fig:published}) reaches $0.06$;
our most faithful runs, on the authors' full data with their settings, read-out and KEGG
proliferation weights, reach $0.16$ (five seeds), still below the walk (Appendix~\ref{app:deviations}).

\subsection{Three data sets, 18 methods}\label{sec:eighteen}

\begin{figure*}[t]
\centering
\includegraphics[width=\textwidth]{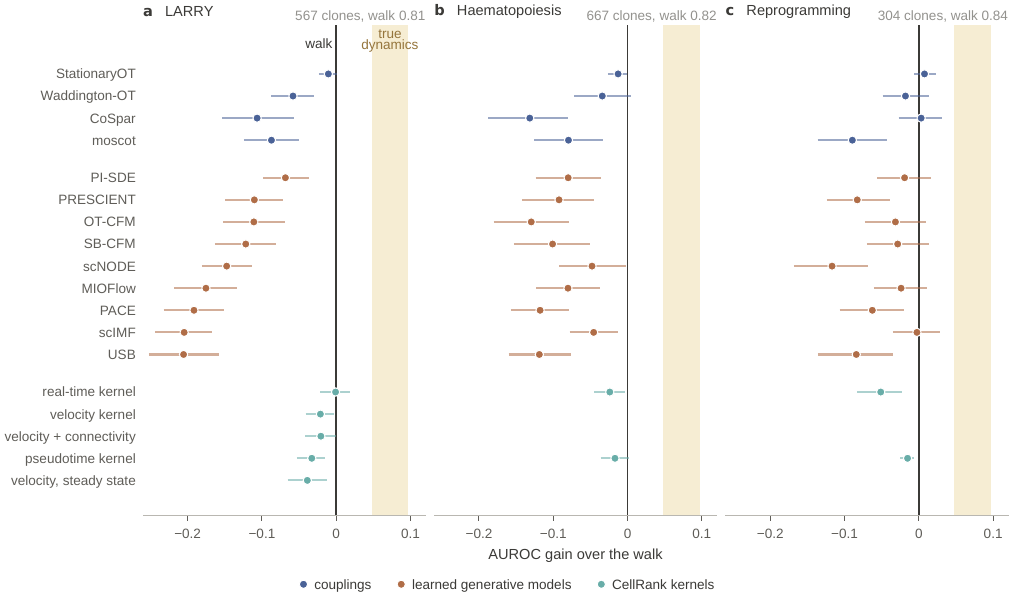}
\caption{Gain over the walk on three lineage-traced data sets. Clone-level AUROC of
each method minus the walk on the same graph and terminal states (dots, 95\% paired bootstrap
over clones), grouped and coloured by method family. Gold band: the gain that true dynamics
carrying fate add in the simulation of Section~\ref{sec:sim}, at the walk's level on these data.
No method is above zero. Missing rows: velocity kernels on data without a splicing layer.}
\label{fig:gain}
\end{figure*}

Figure~\ref{fig:gain} shows every method as a gain over the walk. Of 48 method--data set pairs (18 on LARRY; 15 on each CellTag-multi set,
which has no splicing layer for the velocity kernels),
36 are below the walk, 12 are not separated, and none is above (per-pair 95\% intervals; after a
Holm correction over all 48, 28 remain below). Several sit at zero within their intervals: CellRank's real-time kernel and StationaryOT on LARRY ($-0.001$, $-0.011$), scIMF, CoSpar and
StationaryOT on reprogramming ($-0.003$, $+0.003$, $+0.008$); the farthest is USB on LARRY ($-0.206$). Changing the walk's graph
settings moves it by only $0.010$--$0.021$, less than the distance of most methods from it.
Redefining the terminal states (six definitions on LARRY) keeps it at $0.78$--$0.82$, with the
velocity kernel below it under every definition it can be scored on (Appendix~\ref{app:sens}). Minimum detectable differences range from $0.013$ to $0.077$.

\paragraph{Other scoring choices give the same answer.} With each prediction recalibrated by
cross-fitting, no method has a lower Brier score than the walk in any of the 48 pairs (24 higher,
24 not separated; Appendix~\ref{app:tool}). The read-out of the learned models is not the cause:
a simulated cell whose endpoints reach no scored fate gets $0.5$, but on the clones whose cells all
reach one, none of the 27 learned-model pairs is above the walk (Appendix~\ref{app:sens}).

\paragraph{Six fates instead of two.} Scored over six fates on LARRY (macro one-versus-rest AUROC,
771 clones; walk $0.893$), the learned models stay far below the walk ($-0.13$ to $-0.22$), and the
calibrated score finds its first method above the walk: StationaryOT, $+0.012$ [$0.003$, $0.024$],
above the walk at 12 of its 18 settings and significant after a correction for the 17 methods scored.
Most of the gain comes from one rare fate, megakaryocytes (29 clones; $+0.06$), where the velocity
kernel and the velocity plus connectivity kernel gain as much but lose on neutrophils (Appendix~\ref{app:sens}); the gain is well
below the $0.05$--$0.10$ that true dynamics add in simulation.

\paragraph{Fit and fate rank models differently.} On LARRY, USB and scNODE fit the observed
distributions best (distances $3.7$ and $3.9$ against a floor near $3.6$) and have fate gains of
$-0.206$ and $-0.147$; PI-SDE fits worse ($5.0$) and has the best gain of the learned models ($-0.068$). Fit measures whether the simulated populations match the
observed ones; fate measures whether each cell goes where its sisters went.

\paragraph{Does any method add what the walk lacks?} A method could score below the walk and still
carry fate information the walk does not have. We cross-fit, over clones, a logistic model on the
walk's prediction alone and one on the walk plus the method (standardised logits, 5 folds, 20
repetitions) and take the difference in out-of-fold AUROC with a paired bootstrap. A second kernel
setting used in place of a method adds nothing (at most $0.003$, never separated), as it should. One of 48
pairs adds information: scNODE on CellTag-multi haematopoiesis, $+0.031$ [$0.008$, $0.057$]. It persists
when the clones of the 157 of 889 task cells whose simulated endpoints reach no terminal state are
left out ($+0.048$ [$0.019$, $0.078$]); after a correction for 48 comparisons it is no longer
significant (uncorrected $p \approx 0.01$). No other pair is separated from zero.

\section{Where does the score come from?}

\begin{figure}[t]
\centering
\includegraphics[width=\columnwidth]{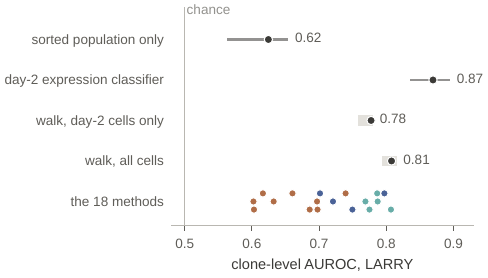}
\caption{Where the score comes from (LARRY, 567 clones). A classifier on day-2 expression alone,
with the sorted population, trained and tested on disjoint clones, outscores the walk; the walk without direction keeps most of
its score when the graph holds day-2 cells only; the 18 methods (coloured by family) sit at or below
the walk. Bars: interval over splits (classifiers) or range over 18 graph settings (walk). Methods coloured
by family as in Table~\ref{tab:setup}.}
\label{fig:source}
\end{figure}

The walk reaches a clone-level AUROC of $0.81$ on LARRY without any direction, so
fate is largely readable before any dynamics are inferred. The signal is how close an early cell
already is to each mature population, and it arrives by two routes: the cell's own expression, and
its connections to later cells on the shared graph (Fig.~\ref{fig:source}).

\paragraph{The cell's own expression.} On LARRY, a classifier given only the day-2 expression of a
cell (and the sorted population it came from), trained and tested on disjoint clones, reaches a
clone-level AUROC of $0.87$; the sorted population alone gives $0.62$. This classifier uses the
clonal labels and is a reference, not a competitor, but it shows that most of the fate signal is
already in the present state.

\paragraph{Later cells on the shared graph.} On LARRY, building the graph from day-2 cells alone,
with the mature cells present on day 2 as terminal states, lowers the walk to $0.76$--$0.78$: part of the signal reaches an early cell through the later
cells and the terminal states it is connected to. Methods and kernel use the same later cells and
the same terminal states, so both receive this part. The walk does not win by this part alone:
built from day-2 cells only, it still exceeds 12 of the 18 methods at its lowest setting ($0.76$),
although they keep the later cells. On CellTag-multi haematopoiesis the early cells alone lose
nothing ($0.830$ versus $0.824$ with all cells): there, fate is written in the early state.

\paragraph{Removing the direction raises the score.} A negative control removes the information about
dynamics step by step while keeping everything else (Table~\ref{tab:ladder}). On LARRY, CellRank's
velocity kernel driven by dynamical velocity scores $0.785$; with the unspliced counts shuffled
across cells, $0.776$; with only the degradation term, $0.739$; with no direction at all (the
connectivity kernel), $0.810$. Removing the direction entirely gives the highest score; the fields
that keep a direction score lower, the decay term lowest. The other rows of the
table run the same ladder against independent measurements of change, where it falls as it should
(Section~\ref{sec:independent}).

\begin{table}[t]
\centering
\small
\setlength{\tabcolsep}{2.5pt}
\caption{Removing the dynamics, step by step. Clonal fate: clone-level AUROC (LARRY). Labelling:
mean cosine with the velocity measured from newly made, metabolically labelled RNA (its production
part, without its decay term; Section~\ref{sec:independent}), gene space. Cell cycle: mean cosine with
the cell-cycle direction measured by FUCCI fluorescence reporters. ``Method'' is the best splicing velocity;
for labelling and cell cycle, ``no direction'' is the average of neighbouring cells' states. Bold: highest in the row.}
\label{tab:ladder}
\footnotesize
\begin{tabular}{@{}lrrrr@{}}
\toprule
 & method & shuffled & decay & no dir. \\
 &  & unspliced & only &  \\
\midrule
Clonal fate, LARRY & $0.785$ & $0.776$ & $0.739$ & $\mathbf{0.810}$ \\
Labelling, haem. & $\mathbf{0.177}$ & $-0.065$ & $-0.636$ & $-0.315$ \\
Labelling, RPE1 & $\mathbf{0.389}$ & $0.022$ & $-0.972$ & $-0.016$ \\
Labelling, organoid & $\mathbf{0.048}$ & $-0.200$ & $-0.464$ & $-0.131$ \\
Cell cycle, RPE1 & $\mathbf{0.439}$ & $0.042$ & $0.113$ & $0.263$ \\
\bottomrule
\end{tabular}
\end{table}

A walk without direction reads both routes. A dynamical method reads them too, and adds its
inferred direction on top; the question is what that direction adds.

\section{Would the test see true dynamics?}\label{sec:sim}

Yes. In a simulation where the true dynamics are known, dynamics that carry fate add $0.05$--$0.10$
AUROC over the walk at the level of the real data sets, large enough for the test to detect (every
pair resolves $0.10$, 18 of 48 resolve $0.05$). In the simulation, falling below the walk takes
directions that favour the wrong fate.

\begin{figure*}[t]
\centering
\includegraphics[width=\textwidth]{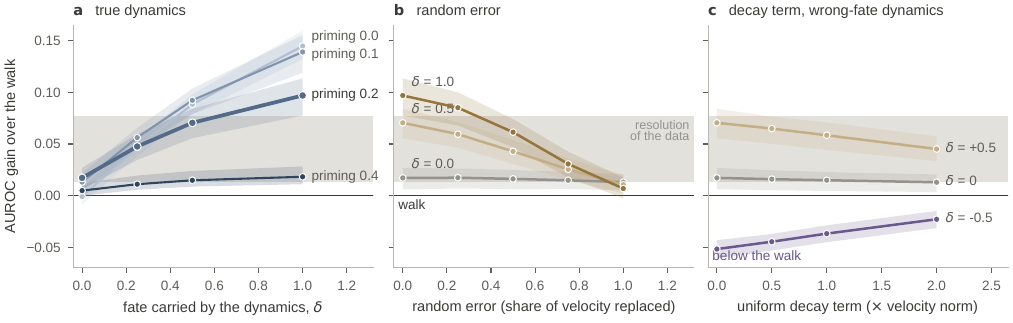}
\caption{Simulated positive control. Gain of CellRank's velocity kernel driven by the true velocity
over the walk on the same graph, on early progenitors of a simulated bifurcation
(means over three seeds; shading, range of the per-seed 95\% intervals). Grey band: minimum
detectable differences on the real data sets. The thick line in (a) is the priming at which the walk
scores as on the real data. (a) The gain grows with how much fate the dynamics carry
($\delta$) and shrinks when the present state already decides fate (priming 0.4). (b) Random error
erodes the gain towards zero and stops there. (c) A uniform decay term shrinks the gain;
only dynamics that favour the wrong fate ($\delta < 0$) fall below the walk.}
\label{fig:mechanism}
\end{figure*}

\begin{figure*}[t]
\centering
\includegraphics[width=\textwidth]{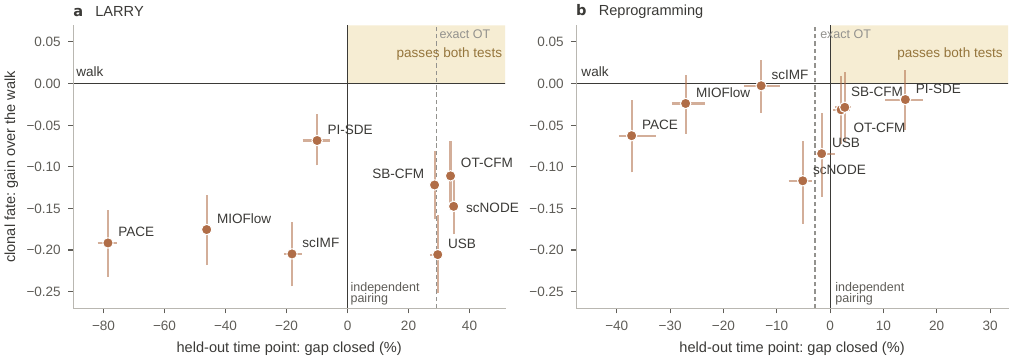}
\caption{The same learned models on two tests. Horizontal: share of the gap from independent
pairing (0\%) to the floor between real batches (100\%) closed on a held-out time point (mean,
range over seeds). Vertical: gain over the walk without direction on clonal fate (clone-level
AUROC, 95\% paired bootstrap). Dashed: exact optimal-transport interpolation. Gold: where a model
that passes both tests would sit; no model is there. On LARRY the models
spread from far below to well above independent pairing, while every one stays below the walk, and
the two orders differ.}
\label{fig:heldout}
\end{figure*}

\paragraph{A simulation with known dynamics.} We simulate a bifurcation with two controls: how much
fate the progenitors' present state carries (priming) and how much the true velocity carries
($\delta$). The true velocity drives CellRank's velocity kernel, which we compare with the
walk on the same graph (Fig.~\ref{fig:mechanism}). The walk follows priming alone
(AUROC $0.49$, $0.67$, $0.83$, $0.98$ at priming $0$, $0.1$, $0.2$, $0.4$). At the level of the real
data sets (priming 0.2, walk $0.83$), true dynamics with $\delta$ from $0.25$ to $1$ add $0.048$--$0.097$. All 48 real pairs have a
minimum detectable difference below $0.097$, and 18 below $0.048$: the positive signal is within
reach. True dynamics lose their room only when the present state nearly decides fate (walk
$0.98$, gain at most $0.02$). A trifurcation, with or without one branch growing faster, gives
$+0.06$ to $+0.12$ at the same walk level (Appendix~\ref{app:sens}).

\paragraph{What takes a method below the walk.} Replacing a growing share of the true velocity with
other cells' velocities erodes the gain towards zero; fully shuffled velocity still stays at or
above the walk, since it points along the common direction of progression. A uniform decay term
also only shrinks the gain. Only dynamics that favour the wrong fate ($\delta = -0.5$) fall below the
walk ($-0.02$ to $-0.05$). Of the 48 real pairs, 36 lie significantly below the walk, by up to
$0.21$. For the velocity kernel the simulation's explanation fits: on LARRY, removing from each
cell's velocity its component between the two fates (16\% of its norm on average) lifts it from
$-0.025$ to $-0.018$, level with the walk (Appendix~\ref{app:sens}).

\paragraph{Would it see a learned model's dynamics?} The velocity kernel above is handed the true
velocity. A model trained on snapshots never sees it: it predicts fate from the present state, and
the most it can reach is a classifier on that state. We ran OT-CFM through the steps used on the real
data on three simulated days (trained across the days, each early cell carried forward, its endpoint
classified by its 20 nearest last-day cells; Appendix~\ref{app:sens}). In the bifurcation a
classifier on the early cells' expression is level with the walk, which already reads all the fate
the state holds, and OT-CFM ends below it in every seed (by $0.04$--$0.07$ on average): couplings
inferred between noisy days carry some cells towards the branch their state does not favour, the
simulation's route below the walk. When one branch turns back so that its terminal cells lie next to
the progenitors primed for the other, the graph links the two and the walk falls to about chance
($0.46$--$0.50$ on average); OT-CFM, which follows the days, rises above it in every seed (by $0.12$
and $0.19$ on average). The test sees a learned model's dynamics where they carry fate the walk
misses. On LARRY that room exists (a classifier on the day-2 state, expression and sorted
population, reaches $0.87$, the walk $0.81$), yet eight of the nine learned models sit further below
the walk than the pipeline's cost,
and the read-out does not explain the gap (Section~\ref{sec:eighteen}).

\section{Which tests do see dynamics?}\label{sec:independent}

Dynamics can be tested. Reported raw, clonal fate cannot tell a method from the walk; a test of
dynamics should be one that
a model without dynamics fails, which means it should ask for what only dynamics give: direction,
rate or timing measured independently of the snapshot. We ran three such tests. Labelling and the
cell cycle test velocity estimates, with the same ladder of controls; the held-out time point tests
the learned models of Sections 3--4 and couplings between days.

\paragraph{Labelled RNA.} Metabolic labelling measures newly made RNA in each cell and so gives a
velocity that does not depend on splicing. On three data sets, splicing velocities separate from
every control without dynamics, in gene space and in the embedding (six of six comparisons;
Table~\ref{tab:ladder}, Appendix~\ref{app:labelling}). The reference is the production part of the
labelling velocity, which we adopted after the first result: the full labelling velocity shares the
decay term with the decay-only control, and against it, in organoid gene space, the decay-only
control scores higher than every splicing velocity (the other five comparisons still separate). Here the ladder falls: shuffling the unspliced
counts, keeping only the decay term or dropping the direction all lower the agreement.

\paragraph{Measured cell cycle.} In RPE1 cells with FUCCI reporters, the cell-cycle direction is
measured by fluorescence. The best splicing velocity aligns with it ($0.439$ mean cosine) well above
the neighbour average ($0.263$) and the other controls.

\paragraph{Held-out time point.} A model trained on the first and last sampling days predicts the
cells of the day in between, the test that TrajectoryNet introduced to machine learning together
with interpolation baselines~\citep{tong2020}. We score each model by the share of the gap it
closes between independent pairing (cells of the two days paired at random and interpolated, no
dynamics, $0\%$) and the floor between two batches of real cells of that day ($100\%$), with the day
held out of training and of the PCA. As a reference, an exact optimal-transport interpolation pairs
the two days by a balanced coupling instead
(Fig.~\ref{fig:heldout}; Table~\ref{tab:heldout}). On LARRY, scNODE, OT-CFM, SB-CFM and USB close $29$--$35\%$, consistently
across seeds and level with an exact optimal-transport interpolation ($29\%$); PI-SDE, scIMF,
MIOFlow and PACE land further from the real day than independent pairing. Across the nine-day gaps of the
reprogramming data PI-SDE helps most ($14\%$); OT-CFM and SB-CFM close $2$--$3\%$. The same models that never beat the walk on clonal
fate separate from the control without dynamics here, and in a different order: scNODE, best here,
has a fate gain of $-0.147$. Waddington-OT's own interpolation, with growth and entropy, closes
$59$--$81\%$ on LARRY and $9$--$12\%$ on reprogramming.

With an independent measurement, velocity estimates, couplings and the learned models separate from
controls without dynamics. For the learned models the gain is at the level of populations, level with
an optimal-transport interpolation on LARRY; what clonal fate could add, a cell-by-cell prediction
beyond the walk, none of them shows yet.

\section{Recommendations}

Agreement with clonal fate can still be informative if it is scored against the walk. When reporting
it:
\begin{enumerate}
\item report the walk on the same graph and terminal states, with its band over graph settings;
\item report the method's gain over the walk with a paired interval over clones and the minimum
detectable difference, and its increment over the walk;
\item state whether the graph and terminal states use later cells, and report the result with early
cells only;
\item choose settings by distribution fit, not by clonal scores.
\end{enumerate}
Our tool takes fate predictions and a clone table and returns the first two. A claim about dynamics is
better supported by a test that a model without dynamics fails: a held-out time point, metabolic
labelling, a measured cell cycle, or the response to a perturbation.

\paragraph{What a better method has to do.} The calibrated score also says where to look. (1)
Start from the walk and add direction: most of the fate signal is already in the present state and
the graph, and methods that fall below the walk lose part of it to directions biased between
fates. (2) The target is concrete: at the real data sets' level, correct dynamics add $0.05$--$0.10$
AUROC, large enough to detect. (3) The room is in early cells whose present state does
not yet decide their fate, and, on LARRY, in megakaryocytes, where the one gain we found sits; where
the present state decides, dynamics have nothing to add. (4) Fit and fate are
different objectives: the model that best reproduces the next day's population is not the one that
sends each cell where its sisters went. A method that aims at fate needs a per-cell objective and
should be selected on both.

\section{Limitations}

The three lineage-traced data sets are haematopoiesis and reprogramming, mostly in vitro; dynamics may
add more in other systems. The simulations (a bifurcation, a trifurcation, fate-dependent growth) are simplified systems and
give orders of magnitude, not exact values. Our PRESCIENT runs score below its published values; we compare with the published values, which
lead to the same conclusion. Each method runs at its published settings, chosen by distribution fit;
settings tuned on clonal fate might score higher, but would use the lineage labels the test holds
out. The positive controls are simulated.

\section{Conclusion}

A high agreement with clonal fate is reachable by a random walk that ignores direction. Only a score
above that walk is evidence of dynamics.

\section*{Software and Data}
The scoring tool, the pipelines for every figure and the data download scripts are available at
\url{https://github.com/xuanruihuang/fatescore}. All data sets are public.

\section*{Impact Statement}
This paper presents work whose goal is to advance the field of Machine Learning. There are many
potential societal consequences of our work, none of which we feel must be specifically highlighted
here.

\IfFileExists{icml2026.bst}{\bibliographystyle{icml2026}}{\bibliographystyle{plainnat}}
\bibliography{main}

\appendix
\onecolumn
\section{Deviations and versions}\label{app:deviations}
\begin{itemize}
\item All four scTimeBench methods run in one environment (torch 2.8, CUDA 12.8): their own pinned
versions have no kernels for the GPU we used.
\item MIOFlow is pinned to commit \texttt{1c6ceb4} (2025-08-12), the last package layout the
scTimeBench runner imports.
\item PI-SDE is trained on the GPU (the benchmark's configuration defaults to the CPU).
\item The scTimeBench generators sample new cells and do not keep cell identity; carrying each
early cell forward is our step.
\item StationaryOT on the reprogramming data uses a looser convergence tolerance on the
marginals ($5.6\times10^{-5}$; the error reached was $3.0\times10^{-5}$).
\item The production part of the labelling reference was added after the first labelling result
was seen; the full reference shares the decay term with the baseline (cosine $0.48$ in the organoid
data).
\item CoSpar~\citep{wang2022cospar} 0.6.0 in its own environment with pinned dependencies (a
current stack broke the neighbour graph); state-only map with its documented defaults (OT start,
smoothing rounds 15, 10, 5, sparsity 0.2, 20 neighbours); fate from its own fate map ('norm-sum').
\item USB~\citep{usb2026} at commit \texttt{0cad29b} (no licence: run for research, not
redistributed) with its Weinreb settings; its growth scale chosen from its own values $\{10, 15\}$ by
fit only. Fate rule fixed before scoring: majority label over a cell's surviving descendants on the
last day, averaged over 10 simulations; the variant without birth and death gives the same verdicts.
\item PACE~\citep{pace2026} at commit \texttt{5f5a10e}, its two documented scRNA presets, the
better fit kept per data set (ipsc on LARRY and reprogramming, op\_cite on haematopoiesis). Its fit
on our data is poor (Table~\ref{tab:fit}); its own evaluation code gives the same distances.
\item PRESCIENT, in three steps, each scored with the published read-out (2,000 trajectories per
cell, 20-neighbour majority label, pseudocount 1, mean over epochs 2,100--2,500). (i) Our first run
on our object (default network, no weights): test $r = 0.06$. (ii) The published LARRY settings from
the authors' repository (two layers of 400 units, noise 0.1, physical days, 50 components, all cells,
2,500 epochs; five seeds): $0.054 \pm 0.010$ without weights. (iii) The authors' full data (GEO
GSM4185642, all 130,887 cells; variable genes with cell-cycle-correlated genes removed, as in their
Methods) and KEGG proliferation weights from the KEGG cell-cycle and apoptosis pathways (mouse;
the authors used the MSigDB versions): $0.082 \pm 0.019$ without and $0.159 \pm 0.017$ with weights
(five seeds); KEGG weights on our object give $0.174 \pm 0.016$. The weights double $r$, as in the
paper ($0.196$ to $0.399$), but from a lower level. On our main task every run is below the
walk (gain $-0.06$ to $-0.16$). Even with the full data, 40\% of
simulated cells reach neither neutrophil nor monocyte. Remaining differences: our re-implementation
of their gene filters and scaling, the pathway gene sets, and the authors' training-cell weights,
which are not public. Fig.~\ref{fig:published} uses the published values ($0.196$ without and
$0.399$ with weights).
\end{itemize}

\section{Distribution fit of the learned models}\label{app:fit}
Settings were chosen by fit alone. Fit is the 2-Wasserstein distance between simulated and observed
cells in the shared 30-component PCA, next to the floor (two batches of the same day) and the
distance of leaving the cells in place. Table~\ref{tab:fit} lists the chosen settings.

\begin{table}[h]
\centering
\small
\caption{Distribution fit (2-Wasserstein distance; lower is better). Two numbers: the two later
days. Ranges: over three seeds (flow matching). In parentheses: early cells (all, not only task cells) whose simulated endpoints reach no
terminal state.}
\label{tab:fit}
\begin{tabular}{llll}
\toprule
 & LARRY (days 4 / 6) & CellTag-multi haem. (day 5) & CellTag-multi reprog. (days 12 / 21) \\
\midrule
Floor (same-day batches) & $\approx 3.6$ & $7.1$ & $5.9$ / $5.4$ \\
No movement & $6.5$ (day 4) & $22.2$ & $10.4$ (day 12) \\
OT-CFM & $3.86$--$3.91$ / $3.94$--$4.20$ & $7.64$--$7.70$ & $6.07$--$6.12$ / $5.95$--$6.21$ \\
SB-CFM & $4.03$ / $4.22$--$4.72$ & $8.10$--$8.25$ & $6.19$--$6.23$ / $6.14$--$6.56$ \\
scNODE & $3.88$ / $3.75$ & $8.7$ (506) & $6.6$ / $6.4$ \\
scIMF & $5.17$ / $5.43$ & $16.4$ & $8.05$ / $6.81$ \\
PI-SDE & $4.95$ / $4.97$ & $10.9$ & $6.36$ / $5.97$ \\
MIOFlow & $5.26$ / $5.41$ (700) & $11.9$ (179) & $7.17$ / $7.17$ \\
USB & $3.72$ / $3.90$ & $8.08$ & $6.29$ / $6.67$ \\
PACE & $5.45$ / $6.66$ & $15.5$ & $8.22$ / $11.1$ \\
\bottomrule
\end{tabular}
\end{table}

scIMF fits CellTag-multi haematopoiesis poorly at its default and at 1,000 epochs ($16.4$ both); we
keep the default and report it.

\section{Scoring tool and statistics}\label{app:tool}
The tool takes, for each early cell, a method's probabilities over the terminal states and a clone
table, and returns: (i) the clone-level AUROC (cell scores averaged within clones) and the cell-level
Pearson $r$; (ii) the connectivity kernel on the same graph and terminal states at 18 settings;
(iii) the gain over the walk with a paired bootstrap over clones (2,000 resamples) and the minimum
detectable difference ($0.714 \times$ interval width, i.e.\ $(z_{0.975} + z_{0.8})/(2 z_{0.975})$);
(iv) the increment over the walk (cross-fitted logistic models over clones). A cell whose
predicted mass on the scored terminal states is zero gets $0.5$. One command reproduces each figure
and table from the released outputs.

\paragraph{Terminal states and simulated topologies.}\label{app:sens} On LARRY we rebuilt the
connectivity kernel and the velocity kernel on the same graph under six terminal-state definitions:
the published annotations ($0.810$ and $0.785$), neutrophil and monocyte only ($0.813$ and
$0.611$), the same on day 6 only ($0.816$ and $0.606$), all mature cells on day 6 only ($0.817$ and
$0.770$), 30 core cells per state ($0.804$ and $0.641$), and CellRank's own macrostates with a fixed
number of states ($0.782$; the velocity kernel finds no monocyte state). The walk stays at
$0.78$--$0.82$ and the velocity kernel is below it under every definition that it can be scored on;
CellRank's automatic choice of the number of macrostates returns a single state on both kernels.
In a trifurcation (macro one-versus-rest AUROC, three seeds), true dynamics add $+0.06$ to $+0.12$
where the walk sits at the real data sets' level ($0.77$--$0.87$), close to the bifurcation's
$0.05$--$0.10$; letting one branch grow twice as fast changes the gains by at most $0.001$; fully
shuffled velocities still add $+0.01$ to $+0.02$, so random error again stops at the walk.

\paragraph{The learned-model pipeline on simulated snapshots.} Three days are sampled from the
simulation of Section~\ref{sec:sim}, 1,500 cells each: early progenitors (the task cells), late
progenitors with the start of both branches, and terminal cells. In the second geometry branch B
turns back after the split and its terminal cells lie behind the stem's start, level with the
progenitors primed for A; the cells in between are not sampled. The walk and the velocity kernel run
on a 30-neighbour graph of all sampled cells (30 components); the state classifier is a cross-fitted
logistic regression on the early cells' expression; OT-CFM uses the model, couplings, integration
and read-out of the real data, with 20,000 training steps and one training seed per run (50,000 and
three on the real data); three simulation seeds each. The classifier and OT-CFM do not depend on $\delta$, which the snapshots do
not contain. Bifurcation, priming $0.1$ / $0.2$ / $0.4$: walk $0.687$ / $0.846$ / $0.982$,
classifier $0.684$ / $0.852$ / $0.984$, OT-CFM $0.652$ / $0.781$ / $0.941$, every seed's interval
below the walk; the true velocity at $\delta = 1$ adds $+0.143$ / $+0.087$ / $+0.013$, as in
Fig.~\ref{fig:mechanism}. Loop, priming $0.2$ / $0.4$: walk $0.498$ / $0.459$, classifier $0.852$ /
$0.984$, OT-CFM $0.616$ / $0.653$ ($+0.09$ to $+0.26$ by seed, every interval above zero; the walk
$0.32$ to $0.62$ by seed). The
velocity kernel with the true velocity varies widely by seed in the loop ($-0.05$ to $+0.34$).

\paragraph{Six fates.} The main LARRY task asks neutrophil or monocyte. We also scored the
probabilities of 17 of the 18 methods (the steady-state velocity kernel's run kept only the two-fate
shares) over six fates (neutrophil, monocyte, basophil, megakaryocyte, mast cell, erythroid;
renormalised over the six; the 771 clones dominated by one of them, with at least three mature
sisters) by macro one-versus-rest AUROC, with no classifier fitted. The walk reaches $0.893$
($0.879$--$0.903$ over the 18 settings). The nine learned models are below it by $0.13$ to $0.22$;
CoSpar, moscot and the pseudotime kernel are also below; Waddington-OT, the real-time kernel and the
two velocity kernels are not separated. StationaryOT is above it: $+0.012$ [$0.003$, $0.024$], above
12 of the 18 settings, bootstrap $p = 0.001$ ($0.009$ after a Holm correction for 17 comparisons). By
fate, most of its gain comes from megakaryocytes (29 clones; $+0.060$ [$0.010$, $0.126$]), the rest
from monocytes and basophils ($+0.011$ each; for monocytes the interval is above zero), against
neutrophils ($-0.009$); over the four fates with at least 30 clones it is
$+0.003$ [$-0.001$, $0.008$]. On megakaryocytes the velocity kernel and the velocity plus
connectivity kernel gain as much ($+0.058$ and $+0.061$, intervals above zero) but lose on
neutrophils ($-0.034$, $-0.028$); the pseudotime kernel does not ($-0.079$). Megakaryocytes are the
one fate where several methods with direction add to the walk; on the other rare fates they do not
(erythroid: velocity kernel $-0.048$, Waddington-OT $-0.052$), and 29 clones are too few to say more.

\paragraph{Clones whose present state leaves fate open.} A logistic classifier on day-2 expression
(50 components, fitted within clone-grouped folds, 5 folds, 10 repetitions) gives each clone an
out-of-fold probability; $|p - 0.5|$ ranks clones by how far their present state decides fate, without
the walk or any method. Within each third we compare every method with the walk on the same clones.
LARRY: the classifier and the walk reach $0.63$ and $0.66$ in the most open third, $0.93$ and $0.90$
in the most decided; no method is above the walk in any third, and the best direction-using
methods do not gain in the open third (real-time kernel $-0.011$, StationaryOT $-0.023$). Reprogramming:
in the open third (102 clones) SB-CFM, OT-CFM and MIOFlow reach $+0.088$ [$-0.019$, $0.199$], $+0.074$
[$-0.036$, $0.189$] and $+0.068$, against $-0.08$ in the middle third. Haematopoiesis: 80\% of clones
are monocyte-dominated, so $|p - 0.5|$ tracks the class prior rather than the cell, and the split is
not informative. A kernel at another graph setting stays at zero in every third.

\paragraph{The walk plus a little direction.} On LARRY we mixed the connectivity kernel with a
direction kernel, $K = (1-w)\,\mathrm{CK} + w\,D$, $w \in \{0.02, 0.05, 0.1, 0.2, 0.3, 0.5\}$, for the
velocity kernel, the pseudotime kernel, the velocity kernel with each cell's velocity component
along the neutrophil-minus-monocyte axis removed (on average 16\% of its norm), a progression-only
velocity kernel, and the real-time kernel. Choosing $w$ on half of the clones and scoring on the
other half (20 splits), the mean gains are $-0.003$, $-0.001$, $-0.003$, $-0.004$ and $+0.014$; no
split has an interval above zero except for the real-time kernel (25\% of splits, $w \approx 0.25$;
$+0.018$ [$0.002$, $0.033$] on all clones). Debiasing lifts the velocity kernel alone from $-0.025$ to
$-0.018$ (not separated) and stops the decline with $w$. On CellTag-multi haematopoiesis the
real-time family gives at most $+0.002$ and a split-validated mean of $0.000$.

\paragraph{Read-out of the learned models.} A learned model's early cell is scored by the fates its
simulated endpoints reach; a cell none of whose endpoints reaches a scored fate gets $0.5$ (on LARRY
101 to 649 task cells per model). Rescoring every learned model on the clones whose task cells all
reach a scored fate, against the walk on the same clones, leaves none of the 27 pairs above the walk;
on LARRY all but PACE (95 clones kept) stay significantly below (gains $-0.03$ to $-0.20$).

\paragraph{Brier score.} On raw probabilities the Brier score also rewards calibration, and the
walk's absorption probabilities are poorly calibrated on CellTag-multi haematopoiesis (Brier
$0.262$), so several methods score lower there. After recalibrating every prediction with a
cross-fitted logistic regression on its logit (5 folds, 20 repetitions), which keeps only the
information in the prediction, no method is below the walk in any of the 48 pairs: 24 are
significantly higher and 24 are not separated (paired bootstrap over clones).

\section{Held-out time point}\label{app:heldout}
Share of the gap closed by every learned model and reference (the data behind
Fig.~\ref{fig:heldout}).

\begin{table}[h]
\centering
\small
\setlength{\tabcolsep}{4pt}
\caption{Held-out time point: share of the gap from independent pairing ($0\%$) to the floor
($100\%$) closed by each model; mean [range] over three seeds (four for the reprogramming references; for scNODE on
reprogramming a fourth seed diverged). LARRY: day 4 from days 2 and 6;
reprogramming: day 12 from days 3 and 21. Negative: further from the real day than independent
pairing.}
\label{tab:heldout}
\begin{tabular}{@{}lrr@{}}
\toprule
 & LARRY & Reprogramming \\
\midrule
scNODE & $35$ [$34$, $36$] & $-5$ [$-8$, $-3$] \\
OT-CFM & $34$ [$34$, $34$] & $2$ [$0$, $4$] \\
SB-CFM & $29$ [$28$, $29$] & $3$ [$1$, $4$] \\
PI-SDE & $-10$ [$-15$, $-6$] & $14$ [$10$, $17$] \\
scIMF & $-18$ [$-21$, $-15$] & $-13$ [$-16$, $-9$] \\
MIOFlow & $-46$ [$-47$, $-45$] & $-27$ [$-30$, $-23$] \\
USB & $30$ [$27$, $31$] & $-2$ [$-3$, $1$] \\
PACE & $-79$ [$-82$, $-76$] & $-37$ [$-40$, $-33$] \\
\midrule
Exact OT interpolation & $29$ [$25$, $33$] & $-3$ [$-4$, $-1$] \\
Last day alone & $6$ [$5$, $6$] & $-46$ [$-48$, $-44$] \\
\bottomrule
\end{tabular}
\end{table}

\section{Metabolic labelling}\label{app:labelling}
Metabolic labelling measures newly made RNA directly and so gives a velocity that does not depend on
splicing. We use scNT-seq haematopoiesis with the labelling velocity published with
dynamo~\citep{qiu2022dynamo} and the scEU-seq RPE1 and intestinal organoid data~\citep{battich2020}.
We compare splicing velocities (scVelo steady-state, stochastic and dynamical models) and
three fields without dynamics (degradation term only, shuffled unspliced counts, neighbour average)
with the production part of the labelling velocity (the labelling velocity minus its own decay
term), by mean cosine over cells in gene space (velocity genes) and after projection into the
embedding. Intervals are 95\% bootstrap over cells. Table~\ref{tab:labelling} lists the best
splicing velocity and the best field without dynamics in each comparison; all six separate.
Against the full labelling velocity, five of the six comparisons still separate (best splicing
velocity against best field without dynamics: haematopoiesis $0.284$ and $0.030$ in gene space,
$0.258$ and $-0.050$ in the embedding; RPE1 $0.207$ and $0.010$, $0.254$ and $-0.011$; organoid
embedding $0.361$ and $0.079$). In organoid gene space the degradation-only field scores higher
($0.114$ versus $0.092$ for the dynamical model; difference $0.023$ [$0.015$, $0.031$]): that reference shares the decay term with the baseline (cosine $0.48$ with
its own decay term), which is why the production part was added (Appendix~\ref{app:deviations}).

\begin{table}[h]
\centering
\small
\caption{Splicing velocities against the production part of the labelling velocity (mean cosine).}
\label{tab:labelling}
\begin{tabular}{lllll}
\toprule
Data set (cells, genes) & Space & Best velocity & Best field without dynamics & Difference [95\% CI] \\
\midrule
scNT-seq haematopoiesis (1,947, 65) & gene & dynamical $0.177$ & shuffled unspliced $-0.065$ & $0.242$ [$0.230$, $0.255$] \\
 & embedding & dynamical $0.253$ & shuffled unspliced $-0.285$ & $0.538$ [$0.493$, $0.584$] \\
scEU-seq RPE1 (2,793, 83) & gene & stochastic $0.389$ & shuffled unspliced $0.022$ & $0.367$ [$0.340$, $0.393$] \\
 & embedding & stochastic $0.332$ & shuffled unspliced $-0.074$ & $0.406$ [$0.363$, $0.444$] \\
scEU-seq organoid (1,373, 157) & gene & dynamical $0.048$ & neighbour average $-0.131$ & $0.180$ [$0.165$, $0.195$] \\
 & embedding & stochastic $0.269$ & degradation only $0.029$ & $0.240$ [$0.203$, $0.278$] \\
\bottomrule
\end{tabular}
\end{table}

\section*{AI disclosure statement}
The author chose the research direction and decided the scope of the work. AI assistants (Claude,
Anthropic) contributed to the experimental design, wrote and ran the analysis code, searched the
literature and drafted the text, under the author's direction. Every number in the paper is produced
by a script from public data and was checked against its source output; each citation was checked
against its publisher record. The author takes responsibility for all content.

\end{document}